\documentclass[preprint,prb]{revtex4}%
\usepackage{amsfonts}
\usepackage{amsmath}
\usepackage{amssymb}
\usepackage{graphicx}%
\providecommand{\U}[1]{\protect\rule{.1in}{.1in}}
\begin{document}
\title{Exact Analytic Second Virial Coefficient for the Lennard-Jones Fluid}
\author{Byung Chan Eu}
\affiliation{Department of Chemistry, McGill University, 801 Sherbrooke St. West, Montreal,
Qc H3A 2K6, Canada}

\begin{abstract}
An exact analytic form for the second virial coefficient, valid for the entire
range of temperature, is presented for the Lennard-Jones fluid in this paper.
It is derived by making variable transformation that gives rise to the
Hamiltonian mimicking a harmonic oscillator-like dynamics. It is given in
terms of parabolic cylinder functions or confluent hypergeometric functions.
Exact limiting laws for the second virial coefficient in the limits of
$T\rightarrow0$ and $T\rightarrow0$ are also deduced for the Lennard-Jones
fluid. They have the forms: $-16\sqrt{2\pi}v_{0}e^{\varepsilon\beta}\left(
\varepsilon\beta\right)  ^{\frac{3}{2}}$ as $T\rightarrow0$ and $4\sqrt
{2}\Gamma\left(  \frac{3}{4}\right)  v_{0}\left(  \varepsilon\beta\right)
^{1/4}$ as $T\rightarrow\infty$, where $\varepsilon$ is the well depth and
$v_{0}=\pi\sigma^{3}/6$ with $\sigma$ denoting the size parameter of the
potential, and $\beta=1/k_{B}T$.

\end{abstract}
\date[Date:
%TCIMACRO{\TeXButton{today}{\today}}%
%BeginExpansion
\today
%EndExpansion
]{}
\startpage{1}
\endpage{102}
\maketitle

\section{Introduction}

The statistical mechanical formula for the second virial coefficient can be
easily computed by using a numerical method or, if temperature is sufficiently
high, by a series expansion method \cite{hirschfelder}. As a matter of fact, a
numerical table is available \cite{hirschfelder} for the Lennard-Jones fluid.
Therefore it does not pose a practical problem, although a low temperature
expansion method is not available. Nevertheless, it would be interesting from
the theoretical and pedagogical as well as aesthetic standpoints and also for
the practical utility, if there were available exact analytic results for
realistic\ interaction potential models, which are valid for the entire range
of temperature. In the literature\cite{garrett,glasser}, an exact analytic
form for the second virial coefficient for the Lennard-Jones (LJ) fluid was
obtained by using coordinate transformation in the cluster integral and
reading off the integral table\cite{ryzhik} to obtain such a form. However,
such transformed integrals have a deeper underlying dynamics that mimics a
dynamical system obeying a harmonic oscillator potential, but reading off an
integral table does not reveal the underlying dynamical structure therein.
Therefore, if such a feature is made evident, one can gain a considerable
insight into the dynamics of the LJ fluid\cite{note1}. The method employed
also provides a valuable lesson on how to handle such integrals that might
appear in the study of statistical mechanics of simple liquids. This method is
not available elsewhere in the literature as far as this author is aware of.

In this paper, we present an exact analytic result for the second virial
coefficient of the Lennard-Jones (LJ) fluid, which is obtained without using
an expansion method and valid for the entire range of temperature. Exact
limiting forms are also deduced therefrom as $T\rightarrow0$ and
$T\rightarrow\infty$. The second virial coefficient obtained is given in terms
of parabolic cylinder functions or confluent hypergeometric functions, which
are convergent and well defined for all values of temperature. The form
presented for the second virial coefficient therefore is valid for all temperatures.

\section{Analytic Solutions}

The second virial coefficient \cite{hill} of the LJ fluid may be written in
the reduced form%
\begin{equation}
B_{2}=-12v_{0}\int_{0}^{\infty}dxx^{2}\left\{  \exp\left[  -4\varepsilon
\beta\left(  x^{-12}-x^{-6}\right)  \right]  -1\right\}  \equiv-12v_{0}I,
\label{1}%
\end{equation}
where $v_{0}=\pi\sigma^{3}/6$, the volume of the contact sphere of diameter
$\sigma$ ($\sigma=$ size parameter of the LJ potential), $\varepsilon$ is the
well depth, and $\beta=1/k_{B}T$, inverse temperature, $k_{B}$ being the
Boltzmann constant. The object of interest is the integral $I$ in Eq.
(\ref{1}). With transformation of variables%
\begin{align}
\alpha &  =\sqrt{\varepsilon\beta},\label{2}\\
y  &  =\frac{4\alpha^{2}}{x^{12}}, \label{3}%
\end{align}
the integral $I$ can be put into the form%
\begin{equation}
I\left(  \alpha\right)  =\frac{\sqrt{2\alpha}}{12}J\left(  \alpha\right)  ,
\label{5}%
\end{equation}
where $J\left(  \alpha\right)  $ is defined by the integral%
\begin{equation}
J\left(  \alpha\right)  =\int_{0}^{\infty}dyy^{-5/4}\left(  e^{-y}%
e^{2\alpha\sqrt{y}}-1\right)  . \label{6}%
\end{equation}
This integral is usually evaluated either by a series expansion method or by a
numerical method. Luckily, this integral is in a form readily available from
the integral table\cite{ryzhik} in a known functional form. However, it can be
analytically evaluated without using an integral table or expansion method, as
will be shown below.

To achieve this aim, perform integration by parts once to obtain $J\left(
\alpha\right)  $ in the form%
\begin{equation}
J\left(  \alpha\right)  =4\left[  \alpha B_{3/4}\left(  \alpha\right)
-B_{1/4}\left(  \alpha\right)  \right]  , \label{7}%
\end{equation}
where $B_{1/4}$ and $B_{3/4}$ are defined by the integrals%
\begin{align}
B_{1/4}\left(  \alpha\right)   &  =\int_{0}^{\infty}dyy^{-1/4}e^{-y}%
e^{2\alpha\sqrt{y}},\label{8a}\\
B_{3/4}\left(  \alpha\right)   &  =\int_{0}^{\infty}dyy^{-3/4}e^{-y}%
e^{2\alpha\sqrt{y}}. \label{8b}%
\end{align}
These integrals are functions of parameter $\alpha$. Differentiating these
integrals with $\alpha$, we obtain a pair of first-order differential
equations%
\begin{align}
\frac{dB_{1/4}}{d\alpha}  &  =\frac{1}{2}B_{3/4}\left(  \alpha\right)
+2\alpha B_{1/4}\left(  \alpha\right)  ,\label{9a}\\
\frac{dB_{3/4}}{d\alpha}  &  =2B_{1/4}\left(  \alpha\right)  . \label{9b}%
\end{align}
This pair of differential equations can be combined to a single homogeneous
second-order differential equation:%
\begin{equation}
\frac{d^{2}\psi}{d\alpha^{2}}-2\alpha\frac{d\psi}{d\alpha}-3\psi=0 \label{10}%
\end{equation}
with the simplified notation%
\begin{equation}
\psi\left(  \alpha\right)  =B_{1/4}\left(  \alpha\right)  . \label{10a}%
\end{equation}
With the transformations%
\begin{equation}
z=\sqrt{2}\alpha\label{11a}%
\end{equation}
and%
\begin{equation}
\psi\left(  z\right)  =e^{\frac{1}{4}z^{2}}\phi\left(  z\right)  \label{11b}%
\end{equation}
the differential equation (\ref{10}) can be transformed into a standard form%
\begin{equation}
\frac{d^{2}\phi}{dz^{2}}-\left(  1+\frac{1}{4}z^{2}\right)  \phi\left(
z\right)  =0. \label{12}%
\end{equation}
This is akin to the Schr\"{o}dinger equation for a harmonic oscillator---of a
negative eigenvalue in the present case. Therefore, it represents a particle
of negative energy subjected to a parabolic potential. It, in fact, is a
differential equation for parabolic cylinder functions
\cite{bateman,abramowitz}. Its two independent solutions, one even and the
other odd function of $z$, are:%
\begin{align}
\phi_{1}\left(  z\right)   &  =e^{-\frac{1}{4}z^{2}}M\left(  \frac{3}{4}%
,\frac{1}{2},\frac{1}{2}z^{2}\right) \nonumber\\
&  =e^{-\frac{1}{4}z^{2}}\sum_{n=0}^{\infty}\frac{\left(  \frac{3}{4}\right)
_{n}}{\left(  \frac{1}{2}\right)  _{n}}\frac{\left(  z^{2}/2\right)  ^{n}}%
{n!},\label{13a}\\
\phi_{2}\left(  z\right)   &  =ze^{-\frac{1}{4}z^{2}}M\left(  \frac{5}%
{4},\frac{3}{2},\frac{1}{2}z^{2}\right) \nonumber\\
&  =ze^{-\frac{1}{4}z^{2}}\sum_{n=0}^{\infty}\frac{\left(  \frac{5}{4}\right)
_{n}}{\left(  \frac{3}{2}\right)  _{n}}\frac{\left(  z^{2}/2\right)  ^{n}}%
{n!}. \label{13b}%
\end{align}
Here $M(a,b,t)$ is a confluent hypergeometric function of Kummer
\cite{abramowitz}:%
\begin{equation}
M\left(  a,b,t\right)  =\sum_{n=0}^{\infty}\frac{\left(  a\right)  _{n}%
}{\left(  b\right)  _{n}}\frac{t^{n}}{n!}, \label{K1}%
\end{equation}
where%
\begin{align}
\left(  a\right)  _{0}  &  =1,\nonumber\\
\left(  a\right)  _{n}  &  =a\left(  a+1\right)  \left(  a+2\right)
\cdots\left(  a+n-1\right)  \;\left(  n\geq1\right)  . \label{K2}%
\end{align}
It is convergent for all values of $t$. Its asymptotic form will be interest
to us later: for positive real $t$ it is given by%
\begin{equation}
M\left(  a,b,t\right)  =\frac{\Gamma\left(  b\right)  }{\Gamma\left(
a\right)  }e^{t}t^{a-b}\left[  \sum_{n=0}^{m-1}\frac{\left(  b-a\right)
_{n}\left(  1-a\right)  _{n}}{n!}t^{-n}+O\left(  t^{-m}\right)  \right]  .
\label{K3}%
\end{equation}
This formula may be used to compute the solutions for a large value of $t$ at
fixed values of $a$ and $b$. The solutions $\phi_{1}\left(  z\right)  $ and
$\phi_{2}\left(  z\right)  $, in fact,\ are parabolic cylinder functions,
which are generic solutions for the Schr\"{o}dinger equations for quadratic
potentials. This implies that the dynamics of the LJ potential fluid closely
resembles that of a harmonic (quadratic) potential.

Therefore the general solution for $\psi(z)$ is:%
\begin{equation}
B_{1/4}(z)=\psi(z)=c_{1}M\left(  \frac{3}{4},\frac{1}{2},\frac{1}{2}%
z^{2}\right)  +c_{2}zM\left(  \frac{5}{4},\frac{3}{2},\frac{1}{2}z^{2}\right)
, \label{14}%
\end{equation}
where $c_{1}$ and $c_{2}$ are constants, which may be determined by
considering the boundary conditions.

Noting that%
\begin{equation}
\frac{d}{dt}M(a,b,t)=\frac{a}{b}M(a+1,b+1,t), \label{15}%
\end{equation}
we find%
\begin{align}
J(\alpha)  &  =4c_{1}\left[  6\alpha^{2}M\left(  \frac{7}{4},\frac{3}%
{2},\alpha^{2}\right)  -\left(  1+4\alpha^{2}\right)  M\left(  \frac{3}%
{4},\frac{1}{2},\alpha^{2}\right)  \right] \nonumber\\
&  +4\sqrt{2}c_{2}\alpha\left[  \frac{10}{3}\alpha^{2}M\left(  \frac{9}%
{4},\frac{5}{2},\alpha^{2}\right)  +\left(  1-4\alpha^{2}\right)  M\left(
\frac{5}{4},\frac{3}{2},\alpha^{2}\right)  \right]  . \label{16}%
\end{align}
The coefficients $c_{1}$ and $c_{2}$ can be determined by examining the
limiting form of $J\left(  \alpha\right)  $ as $\alpha\rightarrow0$ (a
boundary condition). From Eq. (\ref{16})%
\begin{equation}
J(\alpha)=4\left[  -c_{1}+\sqrt{2}c_{2}\alpha+O\left(  \alpha^{2}\right)
\right]  , \label{17a}%
\end{equation}
whereas direct evaluation of $J(\alpha)$ by series expansion of the factor
$\exp\left(  2\alpha\sqrt{y}\right)  $ in Eq. (\ref{6}) yields%
\begin{equation}
J(\alpha)=-4\Gamma\left(  \frac{3}{4}\right)  +2\Gamma\left(  \frac{1}%
{4}\right)  \alpha+O\left(  \alpha^{2}\right)  , \label{17b}%
\end{equation}
where $\Gamma\left(  \frac{1}{4}\right)  $ and $\Gamma\left(  \frac{3}%
{4}\right)  $ are gamma functions: $\Gamma\left(  \frac{1}{4}\right)
=3.62560\cdots$ and $\Gamma\left(  \frac{3}{4}\right)  =1.22541\cdots$.
Comparing\ Eqs. (\ref{17a}) and (\ref{17b}), we find%
\begin{align}
c_{1}  &  =\Gamma\left(  \frac{3}{4}\right)  ,\label{18a}\\
c_{2}  &  =\frac{1}{2\sqrt{2}}\Gamma\left(  \frac{1}{4}\right)  . \label{18b}%
\end{align}
Thus $J(\alpha)$ is now fully determined.

Putting together the results produced up to this point, we finally obtain the
second virial coefficient in the form%
\begin{align}
-B_{2}/v_{0}\sqrt{2}\left(  \varepsilon\beta\right)  ^{1/4}  &  =4\Gamma
\left(  \frac{3}{4}\right)  \left[  6\varepsilon\beta M\left(  \frac{7}%
{4},\frac{3}{2},\varepsilon\beta\right)  -\left(  1+4\varepsilon\beta\right)
M\left(  \frac{3}{4},\frac{1}{2},\varepsilon\beta\right)  \right] \nonumber\\
&  +2\Gamma\left(  \frac{1}{4}\right)  \sqrt{\varepsilon\beta}\left[
\frac{10}{3}\varepsilon\beta M\left(  \frac{9}{4},\frac{5}{2},\varepsilon
\beta\right)  +\left(  1-4\varepsilon\beta\right)  M\left(  \frac{5}{4}%
,\frac{3}{2},\varepsilon\beta\right)  \right]  . \label{19}%
\end{align}
This is the result we have set out to show for the LJ fluid. One may try to
put this result into a simpler form by using the recurrence relations of
Kummer's functions, but the present form appears to be an optimum form.
Rigorous limiting laws can be deduced for $B_{2}$ from Eq. (\ref{19}).

The limiting form of $B_{2}$ as $T\rightarrow\infty$ or $\varepsilon
\beta\rightarrow0$ is easily deduced to be%
\begin{equation}
B_{2}=4\sqrt{2}\Gamma\left(  \frac{3}{4}\right)  v_{0}\left(  \varepsilon
\beta\right)  ^{1/4}\left[  1+O(\varepsilon\beta)\right]  . \label{20a}%
\end{equation}
Thus $B_{2}\rightarrow+0$ as $T\rightarrow\infty$. This means that there is a
high temperature regime where $B_{2}$ is positive, and as $T\rightarrow\infty
$, it vanishes on the positive side according to the law indicated.

On the other hand, the limiting form of $B_{2}$ as $T\rightarrow0$ or
$\varepsilon\beta\rightarrow\infty$ is deduced from the asymptotic forms of
the confluent hypergeometric functions given in Eq. (\ref{K3}). We find
\begin{equation}
B_{2}\left(  T\right)  =-16\sqrt{2\pi}v_{0}e^{\varepsilon\beta}\left(
\varepsilon\beta\right)  ^{\frac{3}{2}}\left[  1+\frac{19}{16\varepsilon\beta
}+\frac{105}{512\left(  \varepsilon\beta\right)  ^{2}}+\cdots\right]  .
\label{21}%
\end{equation}
This limiting law shows that $B_{2}\left(  T\right)  $ tends to negative
infinity according to the formula indicated and is negative below a certain
point in $T$.

These limiting laws for $B_{2}$ are not easily deducible from Eq. (\ref{1}) or
Eq. (\ref{6}) or the series expansion form \cite{hirschfelder} thereof, but
they are simple to deduce if the exact analytic solution presented is made use of.

From the limiting behaviors (\ref{20a}) and (\ref{21}) we can conclude there
must exist a point in $T$ at which $B_{2}\left(  T\right)  $ crosses the $T$
axis (i.e., becomes zero), that is, the Boyle temperature $T_{B}=\beta
_{B}^{-1}/k_{B}$ is defined, as usual, by
\begin{equation}
B_{2}\left(  T_{B}\right)  =0. \label{22a}%
\end{equation}
According to the analytic result obtained, the Boyle point is determined from
a real root of the equation%
\begin{align}
0  &  =4\Gamma\left(  \frac{3}{4}\right)  \left[  6\varepsilon\beta
_{B}M\left(  \frac{7}{4},\frac{3}{2},\varepsilon\beta_{B}\right)  -\left(
1+4\varepsilon\beta_{B}\right)  M\left(  \frac{3}{4},\frac{1}{2}%
,\varepsilon\beta_{B}\right)  \right] \nonumber\\
&  +2\Gamma\left(  \frac{1}{4}\right)  \sqrt{\varepsilon\beta_{B}}\left[
\frac{10}{3}\varepsilon\beta_{B}M\left(  \frac{9}{4},\frac{5}{2}%
,\varepsilon\beta_{B}\right)  +\left(  1-4\varepsilon\beta_{B}\right)
M\left(  \frac{5}{4},\frac{3}{2},\varepsilon\beta_{B}\right)  \right]  .
\label{22}%
\end{align}
Its numerical solution yields
\begin{equation}
T_{B}^{\ast}=\left(  \varepsilon\beta_{B}\right)  ^{-1}=3.41793, \label{23}%
\end{equation}
which should be compared with the literature value \cite{hirschfelder}
$T_{B}^{\ast}=3.42$. This value is practically attained with the truncation of
$M\left(  a,b,\varepsilon\beta_{B}\right)  $ at $n=3$.

In conclusion, we have presented an exact analytic second virial coefficient
of the LJ fluids, which are valid for the entire temperature range, and its
asymptotic behaviors (limiting laws) as $T\rightarrow0$ or $T\rightarrow
\infty$. In view of the agreement of the Boyle temperature with the literature
value deduced from the table for the second virial coefficient
\cite{hirschfelder} it seems to be unnecessary to tabulate the numerical
values of the second virial coefficients; it is rather trivial to do so. The
utility of the result obtained is self-evident for some deductions one can
make about thermodynamic properties of the LJ fluid. Compared to the method
that simply reads off the integral table upon variable transformation in the
integral for $B_{2}$, the present method provides considerable insights into
the dynamics of the LJ liquid.

Finally, it is useful to note that the present exact analytic result for the
second virial coefficient owes its existence to the mathematically favorable
combination of the exponents 12 and 6 of the potential that produces the
closed form for the differential equation for $B_{1/4}\left(  \alpha\right)
$, Eq. (\ref{10}). For other potential models consisting of repulsive and
attractive branches with different exponents we do not obtain a closed
differential equation, but a open hierarchy of first-order differential
equations for integrals making up $J\left(  \alpha\right)  $. The case of
exponents $\left(  9,6\right)  $, namely, the LJ $(9,6)$ potential, produces a
closed inhomogeneous second order differential equation, but its solutions do
not seem to be simple and clean.

The present work was supported in part by the grants from the Natural Sciences
and Engineering Research Council of Canada.

\end{document}